# Improving Performance of Cluster Based Routing Protocol using Cross-Layer Design

Seyed Kazem Jahanbakhsh
Department of Electrical Engineering
Sharif University of Technology
jahanbakhsh@ee.sharif.edu

Marzieh Hajhosseini
Department of Electronic Engineering
Central Tehran branch of Islamic
Azad University
meli_h55@yahoo.com

*Abstract*-The main goal of routing protocol is to efficiency delivers data from source to destination. All routing protocols are the same in this goal, but the way they adopt to achieve it is different, so routing strategy has an egregious role on the performance of an ad hoc network. Most of routing protocols proposed for ad hoc networks have a flat structure. These protocols expand the control overhead packets to discover or maintain a route. On the other hand a number of hierarchical-based routing protocols have been developed, mostly are based on layered design. These protocols improve network performances especially when the network size grows up since details about remote portion of network can be handled in an aggregate manner. Although, there is another approach to design a protocol called cross-layer design. Using this approach information can exchange between different layer of protocol stack, result in optimizing network performances.

In this paper, we intend to exert cross-layer design to optimize Cluster Based Routing Protocol (Cross-CBRP). Using NS-2 network simulator we evaluate rate of cluster head changes, throughput and packet delivery ratio. Comparisons denote that Cross-CBRP has better performances with respect to the original CBRP.

## I. INTRODUCCTION

When utilizing a communication infrastructure is expensive or impossible, mobile users can still communicate with each other through a wireless ad hoc network. Because of limited radio range of mobile nodes a packet is constrained to traverse several hops. Moreover, the mobility of nodes combined with transient nature of wireless links cause network topology changing. Because of these issues a number of routing protocol with different structures created; flat routing protocols and hierarchical routing protocols. In an ad hoc network with flat routing protocol all nodes have the same role in packet forwarding. Therefore protocol performances degrade when the network size increases. In hierarchical routing protocol like fewer nodes have outstanding role in packet routing and other nodes role is inconspicuous.

CBRP is a routing protocol that has a hierarchical-based design [7], [9]. This protocol divides the network area into several smaller areas called cluster. The clustering algorithm of CBRP is Least Cluster Change or LCC [10] means the node with the lowest ID among its neighbors elects as cluster head. Other nodes lie into radio range of this cluster head will be the ordinary nodes of that cluster. Because of mobility of nodes in ad hoc network this is probable that elected cluster head to be too mobile. In addition, because nodes with cluster head role consume more power than ordinary nodes, mobile node with lower ID discharge soon. Through these reasons cluster head election procedure used in CBRP is not suitable.

We used cross-layer design to solve this problem. Although cross-layer approach to network design can increase the design complexity, using a compilation of cross-layer and layered principles to network design in a good approach. In such a structure each layer is characterized by some parameters. These parameters then passed to adjacent layers to help them adapt themselves for best suit the current channel, network, and applications.

To realization this approach, signal strength was used to determine mobility of nodes. This parameter is shared between Phy, MAC and network layers to achieve a better cluster head election algorithm. In fact we used cross-layer approach to elect an appropriate node as cluster head to reduction of cluster head changes rate and therefore superior protocol performances.

## II. RELATED WORKS

A number of clustering algorithms have been proposed in literatures that create clusters that their maximum diameter can be two or more hops. Linked Clustered

Algorithm (LCA) [1], Lowest-ID (LID) [2], Maximum Connectivity (MCC) [3], Least Cluster Change (LCC) and Random Competition Clustering (RCC) [15] are the most famous traditional algorithms. Most of these algorithms have a simple random criterion to elect a cluster head mainly focuses on how to form clusters with a good geographic distribution, such as minimum cluster overlap, etc. These kind of clustering algorithms don't meet stability of clusters; however, it is an important criterion especially when clustering used to support routing. To meet this end some other clustering algorithms was created that considered cluster stability. A number of this kind of clustering algorithms can find in [4], [5], [6], [12] and some other literatures. In [12] cluster head election parameter is node's mobility. In [4] cluster head election is based on mobility and power quantity of nodes. In [5],[6] a weight-based clustering algorithm is proposed. Collection of mobility, link connectivity, power and distance of nodes are gathered to elect a cluster head. The advantage of these algorithms is their precise criterion in cluster head election and therefore more stable clusters creation.

Although there are several proposed clustering algorithms in literatures, a few numbers of them was employed in routing protocols. CBRP and Cluster head Gateway Switch Routing (CGSR) used LCC as their cluster head election algorithm or [15] utilized RCC for this aim.

In this paper we focused on the clustering election of CBRP. We used cross-layer approach to elect cluster heads for it. We replaced MOBIC [12] clustering algorithm instead of its original algorithm means LID.

III. CBRP OVERVIEW

The idea of CBRP [7] is to divide the nodes of an Ad hoc network into a number of overlapping or disjoint clusters. Each cluster elects a node as cluster head. A cluster head exerts *gateway* nodes to communicate with other cluster head through them. In other word a gateway node has at least one cluster head or more. Other nodes in the cluster are *ordinary* nodes. Cluster heads record the membership information for the clusters in two neighbor tables. CBRP's clustering algorithm creates clusters that their diameters are 2 hops. Intra-cluster routes (routes within a cluster) are discovered dynamically using the membership information. CBRP is based on source routing, similar to DSR. This means that inter-cluster routes (routes between clusters) are found by flooding the network with Route Requests (RREQ). The difference is that the cluster structure generally means that the number of nodes disturbed is much less. Flat routing protocols, i.e. only one level of hierarchy, might suffer from excessive overhead when scaled up. Readers is referred to [9] for an analysis of network performance versus scalability. CBRP is fully distributed just like other protocols and this is necessary because of the dynamic essence of Ad hoc network topologies.

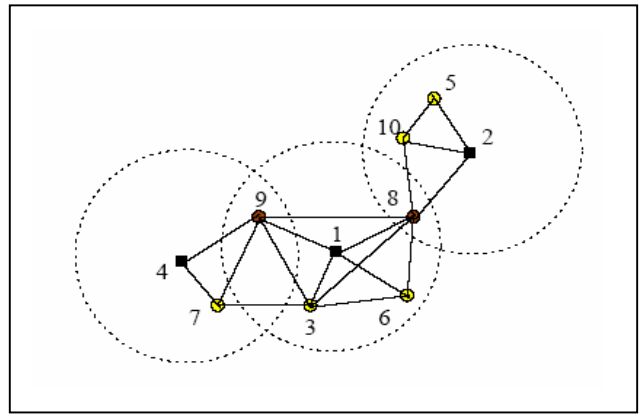

Fig 1. Cluster structure in CBRP

A. *Cluster formation algorithm*

In CBRP, each node transmits some packets named "Hello message" to announce its presence to its neighbor nodes. Upon receiving a hello message, each node updates its neighbor tables. Each node enters the network in the "undecided" state. Every node upon receiving hello message from its neighbors compares its own ID with its neighbor's. If a node distinguishes that its own ID is the lowest ID between its neighbors, this node declares itself as cluster head. Every node that has a bi-directional link to this cluster head will be a member of this cluster [7].

Clusters are identified by their respective cluster heads, which means that the cluster head must change as infrequently as possible. The algorithm is therefore not a strict "lowest-ID" clustering algorithm. A non-cluster head never challenges the status of an existing cluster head. Only when two cluster-heads move next to each other, one of them loses its role as cluster head (LCC)[10]. In Fig.1, node 1 is cluster head for the cluster containing nodes 2, 3, 4 and 5, and node 6 and 8 are cluster heads for two other clusters.

B. *Routing mechanism in CBRP*

Routing in CBRP is based on source routing and the route discovery is done by flooding the network with RREQ. The clustering approach however, means that fewer nodes are disturbed, since only the cluster heads are flooded. If node X seeks a route to node Y, node X will send out a RREQ, with a recorded source route listing only itself initially. Any node forwarding this packet will add its own ID in this RREQ. Each node forwards a RREQ only once and it never forwards it to node that already appears in the recorded route. The source unicasts the RREQ to its cluster head. Each cluster-head unicasts the RREQ to each of its bi-directionally linked neighboring clusters, which has not already appeared in the recorded route through the corresponding gateway. This procedure continues until the target is found or another node can supply the route. When the RREQ reaches the target, the target may choose to memorize the reversed route to the source. It then copies the recorded route to a Route Reply packet and sends it back to the source [7].

In CBRP, a RREQ will always follow a route with the following pattern:
Source → Cluster head → Gateway → Cluster head → Gateway → ⋯ → Destination

## IV. CROSS LAYER APPROACH FOR CBRP: CROSS-CBRP

Mobile Ad hoc networks experience severe topology changes in addition to common problems of other wireless networks. Successive join-and-leave nature of MANET nodes in hierarchical algorithms like CBRP, that is fully dependent on the cluster heads behavior, directly influences the overall network performance. Therefore, wise cluster formation as a mainstream part of these algorithms can improve network performance. In CBRP, cluster formation is performed with a simple and naive approach of the lowest ID. In such a raw selection, every node with the lowest ID between its local neighbors will be cluster head. Obviously, neither the network dynamics nor the clusters stability has been considered. As mentioned earlier, in hierarchical cluster-based MANET, cluster heads play the main role in maintaining the cluster structure and standing against the destructive factors namely mobility. In the cross layer design approach proposed in this paper, cluster formation mechanism and cluster maintenance are considered with respect to proportional mobility of the node towards its neighbors. With this scheme, a node with the lowest mobility and movement in the pre-specified period of time will be named cluster head. By means of cluster head stabilization, network will not suffer from cluster tumbling and local destruction in addition to overheads caused by that.

Recent experimental studies have demonstrated that as the availability of links fluctuates because of channel fading phenomena, the effects of the impairments of the wireless channel on higher-layer protocols are not negligible,. Furthermore, mobility of nodes is not considered. In fact, due to node mobility and node join-and-leave events, the network may be subject to frequent topological reconfigurations. Thus, links and clusters are continuously established and broken. This process in hierarchical cluster-based architecture will result in excessive overhead and cluster head change which degrades performance of the whole network. For the above reasons, new analytical parameters and information from link layer are required to help network layer to determine connectivity conditions; containing mobility and fading channels. In our new approach, the sense of network dynamics and topography changes in physical layer (in the form of received signal power) is fully exploited in network layer cluster formation to achieve energy efficiency and robustness against topological dynamicity [11].

### A. An aggregate local mobility for Cross-CBRP

We use Rayleigh fading model to describe the channel between wireless nodes in a cluster. For a transmitter-receiver separation x, the channel gain is given by:

$$h(x) = L(d_0)(\frac{x}{d_0})^{-n} \xi \quad (1)$$

where $L(d_0) = G_t G_r l^2/16\pi^2 d_0^2$ is the path loss of the close-in distance $d_0$, $G_t$ is the antenna gain of the transmitter, $G_r$ is the antenna gain of the receiver, $l$ is the wavelength of the carrier frequency, $n$ is the path loss exponent ($2 \leq n \leq 6$), and $\xi$ is a normalized random variable that represents the power gain of the fading. Using equation (1), will give us $P_r/P_t \propto x^{-n} \xi$, and by neglecting randomness of fading effect we will have,

$$P_r/P_t \propto x^{-n} \quad (2)$$

The equation (2) shows an inverse n-th power dependence of the radio of received and transmitted power on the physical distance between the transmitter and the receiver.

In reasonably short time scales e.g. a few seconds, the surrounding environment is unlikely to change significantly, therefore .The variable channel gain caused by the effects of multipath, small-scale and large scale fading can be ignored. In this situation the variation of the received signal power will be a good indicator for local mobility of every node.

The ratio of $P_r$ between two successive packet transmissions i.e. periodic "hello" messages from a neighboring node will get us a good knowledge about the relative mobility between two nodes. From this the relative mobility metric $M_Y^{rel}(X)$ at a node Y with respect to X can be define as:

$$M_Y^{rel}(X) = 10 \log_{10}(\frac{p_{r_{X \to Y}}^{new}}{p_{r_{X \to Y}}^{old}}) \quad (3)$$

Now consider a node with $m$ neighbors; there will exist $m$ such values for $M_Y^{rel}(X)$. This situation is depicted in Fig.2. We use the aggregate local mobility value $M_Y$ at any node Y by calculating the variance (with respect to zero) of the entire set of relative mobility samples $M_Y^{rel}(X_i)$, where $X_i$ is a neighbor of Y as proposed in [12]:

$$M_Y = \text{var}\{M_Y^{rel}(X_i)\}_{i=1}^{m} = E[(M_Y^{rel})^2] \quad (4)$$

In this paper $M_Y^{rel}(X_i)$, in (3) is used as a mobility characteristic of a node with respect to its neighbors. As it can be seen from (4) every node is able to calculate $M_Y$, just from a comparison between received powers of "hello" packets in the successive periods of time. Aggregate local mobility of nodes will be included in the advertising packets and broadcasted to neighbors in addition to the node ID and other CBRP's common fields. Resorting to this new field of information, each node makes a table which keeps the track of two parameter for every neighbor; ID and aggregate local mobility. During the cluster formation algorithm, when eligible nodes are competing for taking cluster head role in a distributed manner, aggregate local mobility of every node computed formerly by advertised "hello" packets is compared with aggregate local mobility of its neighbors. For the sake of maximum stability in this heuristic topology control algorithm, the node with the lowest aggregate local mobility will win and take the cluster head role. For better adaptation to uncommon circumstances, lowest ID will be considered just in the rare condition of mobility metric equality. Here, it should be highlighted that $M_Y$ and power estimations which are the building blocks of determinant tables in addition to the tables themselves are gathered, processed and stored locally just with the aid of neighbor's "hello" packets. Despite the fact that each node computes its mobility just with respect to neighbor's contributions independently (an indirect approach), there is no need to have a central node to collect and redistribute node's information with a lot of overhead which means scalability in a mobile Ad hoc network.

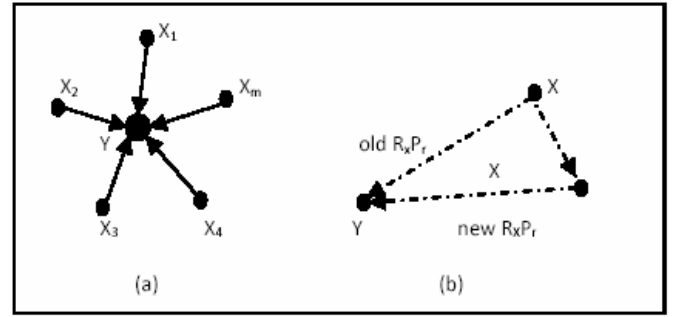

Fig.2 . Calculation of Aggregate Relative Mobility
(a) "Hello" packet reception at Y from neighbor.
(b) Successive Rx Power Measurements at Y due to X.

*B. Distributed Cluster Formation Algorithm for Cross-CBRP*

In order to use the aggregate mobility metric presented in the section "A" for clustering, we propose a two step distributed clustering algorithm which use the mobility metric as a basis for cluster formation. You can find the description of the algorithm in the following paragraph:
All nodes send (receive) "Hello" messages to (from) their neighbors. Each node measures the received power levels of two successive transmissions from each neighbor, and then calculates the pair wise relative mobility metrics using (3). Also, every node extracts the relative mobility metric of every neighbor from received "hello" packet. Then, each node computes the aggregate relative mobility metric $M_Y$ using (4). All nodes start in Cluster-Undecided state. Every node broadcasts its own mobility metric, $M_Y$ (initialized to 0 at the beginning of operation) in a "hello" message to its 1-hop neighbors, once in every Broadcast-Interval (BI) period. If this node is not already in the neighbor table of each neighboring node, will be stored in the neighbor table of them along with a time-out period (TP) seconds as a new neighbor. Otherwise neighboring node situation becomes update. Fig.3 shows the distributed algorithm in details.

This algorithm is distributed. Thus, a node receives the $M_Y$-values from its neighbors, and then compares them with its own. If a node has the lowest value of $M_Y$ amongst

```
A node like m receives a "hello" packet from node n:
m searches its neighbor table
if n is already in the neighbor table
    determine the signal power of the "hello" packet
    received from node n.
    calculate relative mobility metric using (2).
    update neighbor table's fields of node m for n.
    update number of cluster head related to m.
    if m is a cluster head in the proximity of other cluster
    head
    if aggregate relative mobility of m is less than node n
        m remains as cluster head.
        node n give up cluster head role and becomes a
        member of m.
        return.
    else if aggregate local mobility of m is equal to the n
        if m has a lower ID
            m remain as the cluster head.
            n becomes a member of m.
              return.
        else m is a member of its neighbor cluster head
            return.
    else m is a member of its neighbor cluster head
        return.
  else if m is a member and it has no cluster head now
    for every neighbors of m
    if aggregate local mobility of m is less than the
    related neighbor's
    m is cluster head.
    m determines its aggregated relative mobility using (3)
    m broadcasts a "hello" packet to introduce itself to its
    neighbors.
     else m status will change to undecided state.
        a new cluster must be generated.
        return.
    end for
  else record this new neighbor in the neighbor table and
  wait for the next receiving signal of n.
    wait for the next event.
```

Fig. 3. Distributed Cluster Formation Algorithm for Cross-CBRP

all its neighbors, it assumes the status of a cluster head. Then this node broadcasts a "hello" packet to introduce itself as cluster head. In case where the mobility metric of two cluster head nodes is the same, and they are in competition to retain the cluster head status, then the selection of the cluster head is based on the Lowest ID algorithm in which the node with lowest ID gets the status of the cluster head. If a node with cluster member status and with low mobility moves into the range of another cluster head node with higher mobility, re-clustering will not triggered (similar to LCC [10]) because this is in contrary to the network stability and overhead mitigation.

TABLE 1. SIMULATION PARAMETERS

| Parameter | Meaning | Value |
|---|---|---|
| N | Number of Nodes | 100 |
| m × n | Size of the scenario | 1000 x 1000 (m²) |
| Max Speed | Maximum Speed | 10,20,30 (m/s) |
| Tx | Transmission Range | 250 m |
| P.T | Pause Time | 0 sec |

VI. RESULT & DISCUTION

The simulations were performed using the ns-2 network simulator with the MANET extensions [13]. The mobility scenarios were randomly generated using the random waypoint mobility model with input parameters such as maximum speed, number of nodes, area size, etc. Traffic is generated using NS-2 CBR traffic generator. There are simultaneously 60 CBR traffic flows associated with randomly selected disjoint source and destination nodes. Packet size is set to 512 bytes. We used DropTail/PreQueue for implementing the interface queue. This type of queue inserts the routing protocol packets at the head of the queue and all other packets at the back. Size of the queue buffer sets to 50. We implemented Cross-CBRP by doing the required modification on the latest implementation of CBRP in ns-2 environment [14]. The simulation parameters have been listed in Table 1. Two kinds of scenarios were used to evaluate the network performances. Each simulation has been run for 300 seconds, and the results are averaged over 5 randomly generated nodal spatial topologies. We precisely compared performance parameters of our proposed approach with the original CBRP such as rate of cluster head changes, throughput, packet delivery ratio, delay and over head.

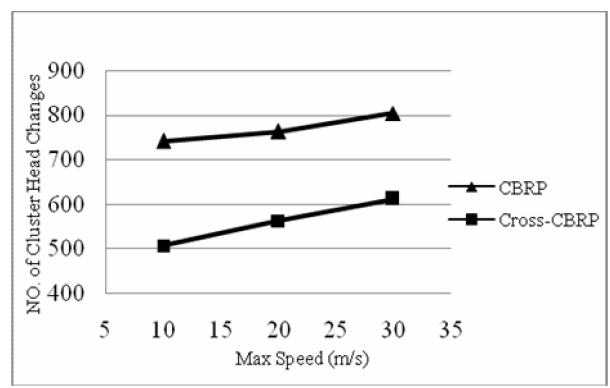

Fig. 4. Number of Cluster Head Changes vs. Speed

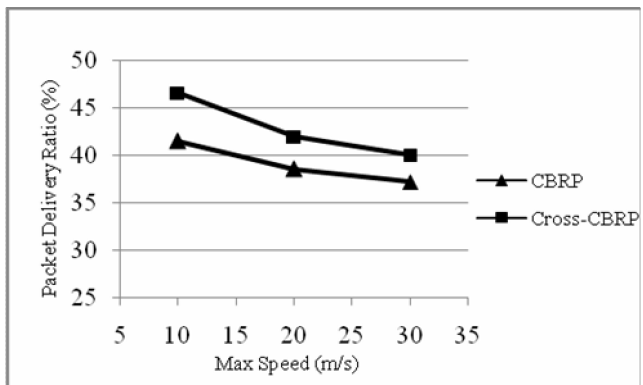
a). Packet Delivery Ratio vs. Speed

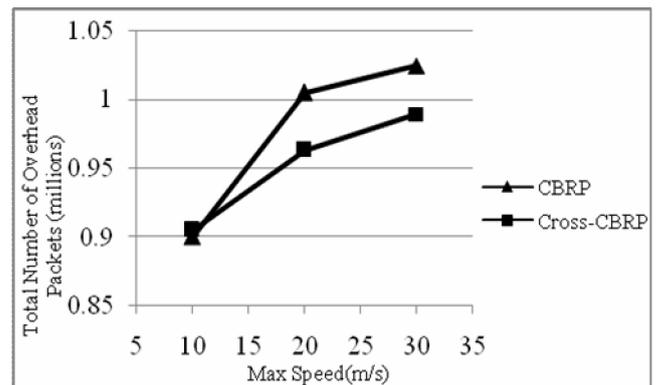
c). Number of Overhead Packets vs. Speed

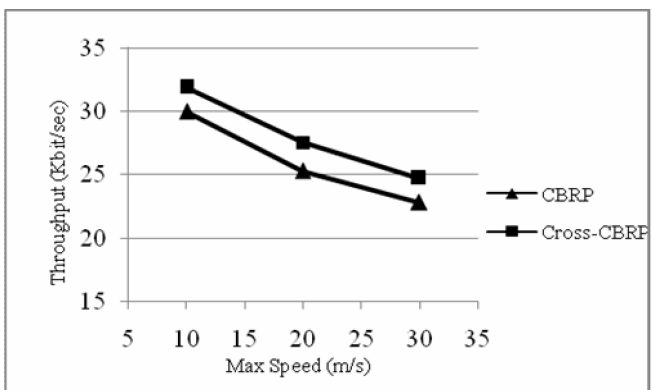
b). Throughput vs. Speed

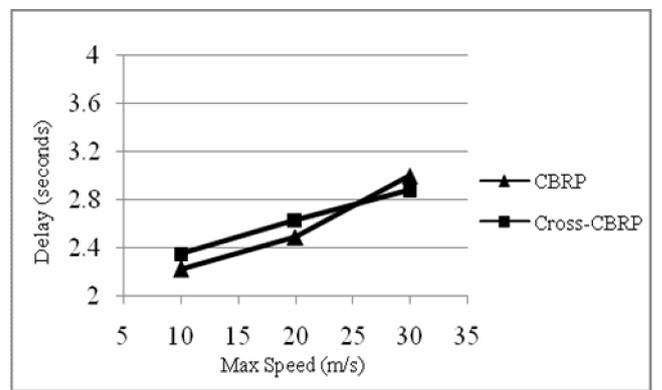
d). Average End-to-end Delay vs. Speed

Fig .5. Performance Comparison between CBRP and Cross-CBRP

Throughput is defined as the average number of data packets received at destinations during simulation time and packet delivery ratio is defined as the total number of data packets sent by traffic sources to the total number of data packets received at destinations, overhead is defined as the total number of control packets including hello packets and finally end-to-end delay is defined as the average time elapsed that a packet originated at the source node, receives at the destination node.

In the first scenario, the rate of packet sent is 4 byte per seconds. Max mobility speed has been considered 10, 20 and 30 m/sec. Fig.4 shows the effect of varying mobility on the performance of Cross-CBRP with respect to CBRP. It can be seen explicitly from Fig.4 that Cross-CBRP outperforms CBRP by averagely 37% improvement for cluster head changes. It is very clear that Cross-CBRP yields a remarkable gain over CBRP because of its capability of adapting itself to the mobility of nodes. From cluster head changes vs. mobility curve, we can conclude that Cross-CBRP is suitable for stable cluster formation in situations involving mobility. Fig.5 (a) demonstrates the packet delivery ratio differences of two algorithms in the existence of mobility. Again we can see that in average the Cross-CBRP performs about 9% better than CBRP because of the cross-layer adaptation technique that has been used in its design. The throughput plays an important role in comparing different network protocols from QoS perspective. Fig.5 (b) demonstrates the results of measured throughput for two previously discussed protocols. The performance results show more efficient behavior of Cross-CBRP in comparison with CBRP with respect to mobility. As it is apparent from the Fig.5 (b), the Cross-CBRP outperforms CBRP about 8.5% which again supports this claim that increasing cluster stability we will give us better network performance.

The total number of control packets as the protocol overhead of these two protocols is compared with each other in (c) .As depicted from this figure it can be seen that Cross-CBRP performs better than CBRP according to this fact that it decreases the cluster reformations. Finally in Fig.5 (d) the end-to-end delay of two protocols analyzed which demonstrates an ignorable difference

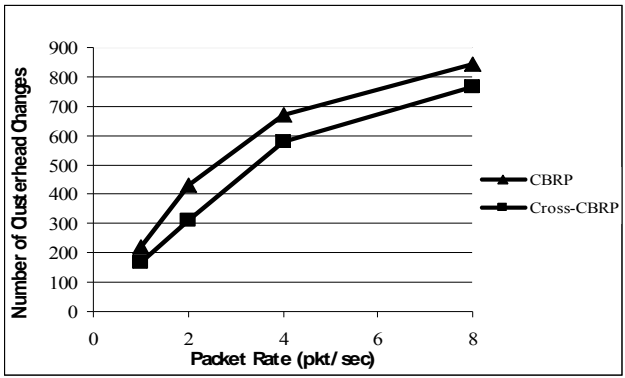

Fig.6. Number of Cluster Head Changes vs. packet Rate

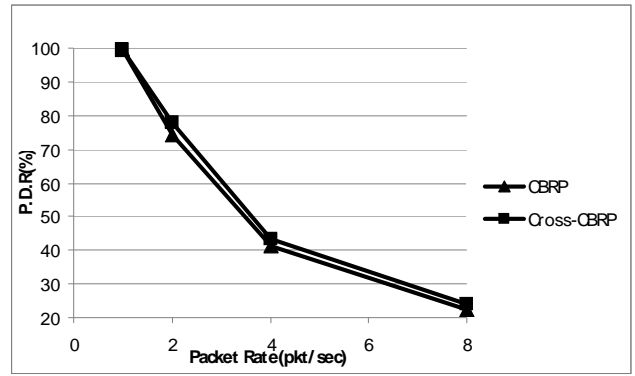

Fig.7. Packet Delivery Ratio vs. Packet Rate

between them.

In the second scenario we changed the sent packet rate from 1 pkt/sec to 8 pkts/sec. In this scenario we intend to study effect of varying traffic on the performance of Cross-CBRP with respect to original-CBRP. As shown in Fig.6 cluster head change rate increases when the packet rate is augmented. When injected traffic to network increases some reasons can cause packets do not received by down stream node - for example lack of route or impossibility to access to the media – so packets will hold in interface queue. If this buffer overflows the last incoming packet will discard. Therefore, if there are some hello packet in this queue these hello packets reach to the neighbors nodes by delay. Two cluster head may have a uni-directional link with each other in this elapsed time; so both of them remain as cluster head until their link changes to bi-directional link. When hello messages reach to destination uni-directional link can change to bi-directional. Therefore, one of adjacent cluster heads must change its role. This will cause the cluster head changing rate increase by increasing injected traffic to network. As we seen in this figure the rate of cluster head changes in Cross-CBRP is out perform original CBRP about 30% in low packet rate and 10% in a high packet rate. This is again because of Cross-CBRP capability to adapting itself to the network conditions.

Fig.7 shows the packet delivery ratio versus packet rate. Injecting traffic to the network causes degrading probability of access to the media. Whatever, the injected packet to network increases, packet delivery ratio decreases. Again we can see that in average the Cross-CBRP performs about 9% better than CBRP because of the cross-layer adaptation technique that has been used in its design. The last graph is related to network throughput.

Fig.8 shows the results of measured throughput for two previously discussed protocols. Although packet delivery ratio decreases when traffic rate increases, increasing throughput continued. This is because of increasing amount of injected traffic in the network that cause number of received packet bytes increases. Again it can be seen from Fig.8 that Cross-CBRP throughput outperforms original-CBRP about 10% when traffic increases.

## VII. CONCLUSION & FUTURE WORKS

Clustering algorithms as discussed in section IV provide more efficient way to utilize the network resources like bandwidth and energy. Mobility of nodes in MANET has a destructive role in the efficient resource allocation. In this paper, we presented a new approach to cross-layer design of CBRP to enhance its efficiency with respect to the existence of mobility in Ad hoc networks. Cross-CBRP, by considering multiple layers such as physical, MAC and network layer tries to provide an adaptive clustering algorithm. Using ns-2 we demonstrated that Cross-CBRP outperforms CBRP in different performance factors. Therefore, we conclude that Cross-CBRP, using mobility parameter sensed via physical layer, is able to behave much better than CBRP which does not account for mobility issues at all. We believe that the cross-layer approach for designing clustering protocol for Ad hoc and wireless sensor networks is a productive field of research. It is possible, to account for other parameters from the physical layer such as channel state to provide more reliable adaptive clustering protocols regarding varying behavior of wireless channels like fading and noise effects.

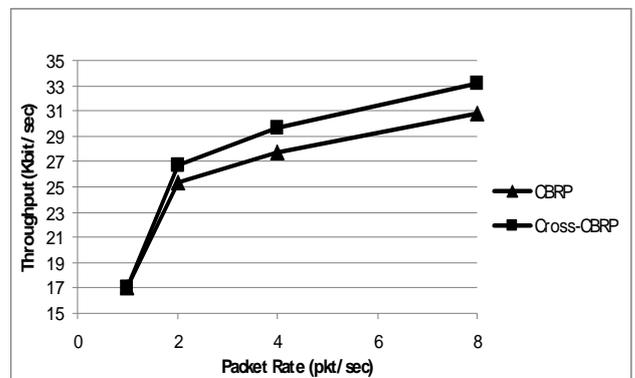

Fig.8. Throughput vs. Packet Rate